\documentclass[11pt,twoside]{article}
\usepackage{bm}
\usepackage{latexsym}
\usepackage{dcolumn}
\usepackage{amsfonts,amssymb}
\usepackage{graphicx}
\usepackage{epsfig}
\usepackage{psfrag}
\usepackage{nccmath}
\usepackage{moresize}
\usepackage{enumerate}
\usepackage[hang,flushmargin]{footmisc}
\usepackage[titletoc,toc]{appendix}
\usepackage{lipsum}
\usepackage{subcaption}
\usepackage{hyperref}
\usepackage{rotating}
\usepackage{multirow}
\usepackage{mathtools}
\usepackage{stackengine}   

\usepackage{cite}
\usepackage{tikz}
\usetikzlibrary{positioning}

\usepackage{stackengine}
\usepackage{scalerel}

\stackMath

\newcommand{\be}{\begin{equation}}
\newcommand{\ee}{\end{equation}}
\newcommand{\bea}{\begin{eqnarray}}
\newcommand{\eea}{\end{eqnarray}}
\newcommand{\bse}{\begin{subequations}}
\newcommand{\ese}{\end{subequations}}
\newcommand{\bce}{\begin{center}}
\newcommand{\ece}{\end{center}}
\newcommand{\bfg}{\begin{figure}}
\newcommand{\efg}{\end{figure}}
\newcommand{\bit}{\begin{itemize}}
\newcommand{\eit}{\end{itemize}}
\newcommand{\bed}{\begin{description}}
\newcommand{\eed}{\end{description}}
\newcommand{\ben}{\begin{enumerate}}
\newcommand{\een}{\end{enumerate}}
\newcommand{\nn}{\nonumber}

\newcommand{\pa}{\partial}
\newcommand{\fr}{\frac}
\newcommand{\sq}{\sqrt}
\newcommand{\no}{\noindent}

\def\a  {\alpha}
\def\b  {\beta}
\def\c  {\gamma}
\def\C  {\Gamma}
\def\d  {\delta}

\def\e  {\epsilon}

\def\f  {\phi}
\def\F  {\Phi}
\def\k  {\kappa}

\def\L  {\Lambda}
\def\m  {\mu}
\def\n  {\nu}

\def\O  {\Omega}

\def\r  {\rho}
\def\th {\theta}
\def\s  {\sigma}

\def\th  {\theta}
\def\vph {\varphi}

\def\le {\left}
\def\ri {\right}
\newcommand{\cA}{\mathcal A}

\newcommand{\cH}{\mathcal H}

\newcommand{\cL}{\mathcal L}

\newcommand{\cO}{\mathcal O}

\newcommand{\cQ}{\mathcal Q}

\newcommand{\cS}{\mathcal S}
\newcommand{\cT}{\mathcal T}

\newcommand{\fw}{\mathfrak w}
\newcommand{\nab}{\nabla\!}
\newcommand{\nt}{\widetilde{\nabla}\!}

\newcommand{\Rt}{\widetilde{R}}
\newcommand{\st}{\widetilde{s}}
\newcommand{\Ct}{\widetilde{\C}}

\newcommand{\hg}{\widehat{g}}
\newcommand{\hl}{\widehat{\lambda}}

\newcommand{\hR}{\widehat{R}}
\newcommand{\hS}{\widehat{\cS}}
\newcommand{\hLm}{\widehat{\cL}^{(m)}}

\newcommand{\keff}{\k_{\text{\scriptsize eff}}}

\newcommand{\dmt}{\d^{(m)}}
\newcommand{\dmp}{\d^{(m)}_{_0}}
\newcommand{\rmt}{\r^{(m)}}

\newcommand{\tmt}{\th^{(m)}}

\newcommand{\Om}{\O^{(m)}}

\newcommand{\Omp}{\O^{(m)}_{_0}}

\newcommand{\sw}{\mathsf w}

\newcommand{\Hp}{H_{_0}}
\newcommand{\fp}{\f_{_0}}
\newcommand{\tp}{t_{_0}}

\newcommand{\se}{\s^{(8)}}
\newcommand{\sep}{\s^{(8)}_{_0}}
\newcommand{\cp}{\c_{_0}}
\newcommand{\cpp}{\c _{_1}}
\newcommand{\gfp}{f_{_0}}

\newcommand{\bdm}{\begin{displaymath}}
\newcommand{\edm}{\end{displaymath}}

\long\def\symbolfootnote[#1]#2{\begingroup%
	\def\thefootnote{\fnsymbol{footnote}}\footnote[#1]{#2}\endgroup}
\begin{document}

\markboth{Mohit Kumar Sharma and Sourav Sur}
{Growth of Matter Perturbations in Scalar coupled Metric-Torsion Cosmology}

%\catchline{}{}{}{}{}
%
%%%%%%%%%%%%%%%%%%%%%%%%%%%%%%%%%%%%%%%%%%%%%%%%%%%%%%%%%%%%%%%%%%%%

\title{\LARGE{Growth of matter perturbations in an interacting dark energy scenario emerging from metric-scalar-torsion couplings}}

\author{Mohit Kumar Sharma\footnote{email: mr.mohit254@gmail.com}~ and~
Sourav Sur\footnote{email: sourav@physics.du.ac.in, sourav.sur@gmail.com}
 \\ \\
{\normalsize \em Department of Physics \& Astrophysics}\\
{\normalsize \em University of Delhi, New Delhi - 110 007, India}
}

\date{}

\maketitle

\begin{abstract}
We study the growth of linear matter density perturbations in a modified gravity approach of scalar field couplings with metric and torsion. In the equivalent scalar-tensor formulation, the matter fields in the Einstein frame interact as usual with an effective dark energy component, whose dynamics is presumably governed by a scalar field that sources a torsion mode. As a consequence, the matter density ceases to be self-conserved, thereby making an impact not only on the background cosmological evolution but also on the perturbative spectrum of the local inhomogeneities. In order to estimate the effect on the growth of the linear matter perturbations, with the least possible alteration of the standard parametric form of the growth factor, we resort to a suitable Taylor expansion of the corresponding exponent, known as the growth index, about the value of the cosmic scale factor at the present epoch. In particular, we obtain an appropriate fitting formula for the growth index in terms of the coupling function and the matter density parameter. While the overall parametric formulation of the growth factor is found to fit well with the latest redshift-space-distortion (RSD) and the observational Hubble (OH) data at low redshifts, the fitting formula enables us to constrain the growth index to well within the concordant cosmological limits, thus ensuring the viability of the formalism. 
\end{abstract}

\vspace{10pt}
\no

%\maketitle

%\tableofcontents

%%%%%%%%%%%%%%%%%%%%%%%%%%%%%%%%%%%%%%%%%%%%%%%%%%%%%%%

\section{Introduction}

The effect of the evolving dark energy (DE) on the
rate of the large-scale structure (LSS) formation has been
a prime area of investigation in modern cosmology, particularly
from the point of view of asserting the characteristics of the 
respective DE component
\cite{CST-rev,FTH-rev,AT-book,wols-ed,MCGM-ed}.
While the observations grossly favour such a component to be
a cosmological constant $\L$
\cite{wmap9-CP,wmap9-Fin,bet-SN,Planck15-CP,Planck15-DEMG,scol-SN,Planck18-CP},
a stringent {\em fine-tuning} problem associated with the corresponding
model, viz. $\L$CDM (where CDM stands for cold dark matter), has prompted 
extensive explorations of a dynamically evolving DE from various perspectives. 
Moreover, certain observational results do provide some scope of a plausible
dynamical DE evolution, albeit upto a significant degree of mildness. In this 
context, it is worth noting that however mild the DE dynamics may be, at 
the standard Friedmann-Robertson-Walker (FRW) background cosmological level, 
there may be substantial effects of such dynamics on the spectrum of the 
linear matter density perturbations. Hence, the analysis of the observational 
data on the evolution of such perturbations, or the LSS growth data, is crucial
for constraining dynamical DE models of all sort.

Apart from the commonly known dynamical DE models involving
scalar fields (such as quintessence, kessence, and so on
\cite{CRS-quin,CLW-quin,tsuj-quin,AMS2000-kess,AMS2001-kess,
MCLT-kess,schr-kess,SSSD-kess,SS-dquin}),
a considerable interest has developed in recent years on the
cosmological scenarios emerging from scalar-tensor equivalent modified gravity 
(MG) theories
\cite{NO-mgDE,tsuj-mgDE,CFPS-mg,papa-ed,NOO-mg}
that stretch beyond the standard principles of General Relativity (GR). Such 
scenarios are particularly useful for providing plausible resolutions to the 
issue of cosmic {\em coincidence} which one usually encounters in scalar field DE
models and in the concordant $\L$CDM model. One resolution of course comes from
the consideration of plausible contact interaction(s) between a scalar field 
induced DE component and the matter field(s)
\cite{CST-rev,AT-book,WS-intDE,amend-intDE,CPR-intDE,FP-intDE,
CW-intDE,CHOP-intDE,amend-rev,BBM-uDE,BBPP-uDE,GNP-uDE,
FFKB-uDE,CMV-mm,MV-mm,NOO-mfR,
SVM-mm,LMNV-mdhost,CM-mmg1,CM-mmg2,CMR-maf,CDS-MMT,SDC-MMT},
which the scalar-tensor formulations naturally lead to, under conformal 
transformations
\cite{FM-ST,frni-ST,BP-ST,BEPS-ST,TUMTY-ST,ENO-ST,CHL-ST,BGP-ST,ST-ST,ST-KT}.
A DE-matter (DEM) interaction makes the background matter density $\rmt(z)$ drifting 
from its usual ({\em dust}-like) evolution with redshift $z$, thereby affecting the 
drag force on the matter perturbations. The evolution of the matter density contrast
$\,\dmt(z) := \d\rmt(z)/\rmt(z) \,$ and the growth factor $f(z)$ of the matter 
perturbations are therefore not similar to those in the non-interacting models, in 
which the field perturbations decay out in the sub-horizon regime, while 
oscillating about a vanishing mean value. Actually, the decaying nature persists 
in the interacting scenarios as well, however with the oscillations about a value 
proportional to the amount of the interaction, measured by the strength of the scalar 
field and matter coupling. As such, the field perturbations contribute to the velocity 
divergences of the matter, affecting in turn the evolution of $\dmt(z)$
\cite{AT-book,koiv-grow}.
Strikingly enough, a DEM interaction can make the growth factor $f(z)$ acquiring a 
value $> 1$ at large $z$, which necessitates the modifications of the commonly known 
$f(z)$ parametrizations in the literature
\cite{PA-grow,PG-grow,GP-grow,WYF-grwpara,PAB-grwpara,BBS-grwpara},
such as the well-known parametrization $f(z) = \le[\Om(z)\ri]^{\c(z)}$, where $\Om(z)$ 
is the matter density parameter and $\c(z)$ is the so-called {\em growth index}
\cite{PG-grow,GP-grow,WYF-grwpara,PAB-grwpara,BBS-grwpara,BP-grwindx,SBM-grwindx,bat-grwindx,MBMDR-grwindx,
PSG-grwindx,BA-grwindx,ABN-naDEp,BNP-fR}.
Our objective in this paper is to attempt such a modification and demonstrate its 
utilization in constraining a DEM scenario emerging from a typical scalar-tensor 
equivalent `geometric' alternative of GR, viz. the metric-scalar-torsion (MST) 
cosmological theory, formulated recently by one of us (SS)
\cite{SSASB-MST,ASBSS-MSTda,ASBSS-MSTpp},
on the basis of certain considerations drawn from robust argumentations that have 
been prevailing for a long time 
\cite{BOS-MST,BOS-QGbook,NFS-MST,FRM-MST,shap-trev}.

MST essentially forms a class of modified (or `alternative') gravity theories that 
contemplates on the appropriate gravitational coupling(s) with scalar field(s) in 
the Riemann-Cartan ($U_4$) space-time geometry, endowed with curvature as well as 
{\em torsion}. The latter being an inherent aspect of a general metric-compatible 
affine connection, is considered as the entity that naturally extends the geometric 
principles of GR, not only from a classical viewpoint, but also from the perspective
a plausible low energy manifestation of a fundamental (quantum gravitational) 
theory\footnote{See the hefty literature on the vast course of development of the 
torsion gravity theories in various contexts, the physical implications and 
observable effects of torsion thus anticipated, as well as searched extensively 
over several decades
\cite{shap-trev,einst-rel,traut,HVKN-trev,akr-tbook,SG-tbook,SS-tbook,HO-trev,
blag-book,CL-extgrav,popl-trev,WZ-trev,CLS-tclass,bloom-tmax,tsam-tmax,
GS-tmax1,GS-tmax2,SSASB-tmax,pmssg,ham,ssgss-kr1,skpmssgas-kr,skpmssgss-kr,
skssgss-kr,ssgss-kr2,dmssgss-kr1,sssdssg-kr,CMS-kr,bmssssg-kr1,ssgss-kr3,
dmssgss-kr2,bmssssg-kr2,BC-kr,adbmssg-kr,HRR-proptor,CF-proptor,saa-proptor,
BS-proptor,HMS-pvtor,bmssgss-pvtor,bmssssgss-pvtor,dmpmssg-pvtor,FMT-pvtor,
mer-Holst,ban-Holst,ST-Holst,HORB-skew,ROH-skew,Ni-skew,YN-PGT,SNY-PGT,
BHN-PGT,HB-PGT,LC-PGT,NRR-PGT,BF-fT,LSB-fT,CCDDS-fT,GLSW-fT,CCLS-fT,BCFN-fT,
VFS-sqtor,VCVM-sqtor,KS-deg1,FR-texpt,KRT-texpt,BF-texpt,HOP-texpt,CCR-texpt,
CCSZ-texpt,LP-texpt}.
}. 
Nevertheless, conventional $U_4$ theories (of Einstein-Cartan type) are faced with 
a stringent uniqueness problem while taking the minimal couplings with scalar fields 
into consideration
\cite{BOS-MST,BOS-QGbook,NFS-MST,FRM-MST,shap-trev}.
Such couplings are simply not conducive to any unambiguous assertion of equivalent 
Lagrangians upon eliminating boundary terms in the usual manner. The obvious wayout 
is the consideration of explicit non-minimal (or, contact) couplings of the scalar 
field(s) with, most appropriately, the entire $U_4$ Lagrangian given by the $U_4$ 
curvature scalar $\Rt$
\cite{SSASB-MST}.
For any particular non-minimal coupling of a scalar field $\f$ with $\Rt$, the 
resulting (MST) action turns out to be equivalent to the scalar-tensor action, as 
the trace mode of torsion, $\cT_\m\,$, gets sourced by the field $\f$, by virtue 
of the corresponding (auxiliary) equation of motion. On the other hand, torsion's 
axial (or, pseudo-trace) mode $\cA_\m$ can lead to an effective potential, for e.g. 
a mass term $\, m^2 \f^2$ (with $m =$ constant) in that scalar-tensor equivalent 
action, upon implementing a norm-fixing constraint ($\cA_\m \cA^\m =$ constant) as 
in the Einstein-aether theories
\cite{JM-EA-2001,JM-EA-2004,JLM-EA},
or incorporating a $\f$-coupled higher order term $(\cA_\m \cA^\m)^2$
\cite{SSASB-MST}.
Such a mass term is shown to play a crucial role in giving rise to a viable 
cosmological scenario marked by a $\f$-induced DE component with a weak enough
dynamical evolution amounting to cosmological parametric estimations well within 
the corresponding observational error limits for $\L$CDM. This also corroborates 
to the local gravitational bounds on the effective Brans-Dicke (BD) parameter 
$\fw$, which turns out to be linear in the inverse of the MST coupling parameter 
$\b$
\cite{SSASB-MST}.

Particularly intriguing is the MST cosmological scenario that emerges under a 
conformal transformation from the Jordan frame to the Einstein frame, in which 
the effective DE component interacts with the cosmological matter ({\it a priori} 
in the form of {\it dust}). Nevertheless, the crude estimate of $\b$ (or of the 
parameter $s = 2\b$, that appears in the exact solution of the Friedmann equations), 
obtained under the demand of a small deviation from the background $\L$CDM evolution
\cite{SSASB-MST},
requires a robust reconciliation at the perturbative level. On the other hand,
the methodology adopted here can in principle apply to any scalar-tensor cosmological 
scenario, once we resort to the dynamics in the Einstein frame.

Now, the methodology of our analysis purports to fulfill our objective mentioned 
above. Specifically, we take the following course, and organize this paper 
accordingly: in section \ref{sec:MST}, we review the basic tenets of MST cosmology
in the standard FRW framework, and in particular, the exact solution of the
cosmological equations in the Einstein frame that describes a typical interacting
DE evolution. Then in the initial part of section \ref{sec:MPGrowth}, we obtain the 
differential equations for $\dmt(z)$ and $f(z)$, and get their evolution profiles 
by numerically solving those equations for certain fiducial settings of the 
parameters $s = 2\b$ and $\, \Omp \equiv \Om\big\vert_{z=0}$. Thereafter, in 
subsection \ref{sec:GF}, we resort to a suitable growth factor parametrization, 
demanding that an appropriate expansion of the growth index $\c(z)$ about the 
present epoch ($z = 0$) should adhere to the observational constraints on the 
growth history predictions at least upto $z \simeq 1$ or so. Next, in subsection 
\ref{sec:GIFit}, we attain the pre-requisites for the growth data fitting with 
the quantity $f\se(z)$, where $f(z$) is as given by its chosen parametrization, 
and $\se(z)$ is the root-mean-square amplitude of matter perturbations within a 
sphere of radius $8\,$Mph$^{-1}$. Finally, in section \ref{sec:RSDEst}, we estimate 
the requisite parameters $s$, $\Omp$ and $\, \sep \equiv \se\big\vert_{z=0}\,$, and 
hence constrain the model by fitting $f\se(z)$ with a refined sub-sample of the
redshift-space-distorsion (RSD) data, and its combination with the observational 
Hubble data
\cite{RDR-bao}. 
In section \ref{sec:Concl}, we conclude with a summary of the work, and an account 
on some open issues.

\bigskip
\no 
{\large \sl Conventions and Notations}: We use metric signature $\, (-,+,+,+)$ 
and natural units (with the speed of light $c = 1$), and denote the metric
determinant by $g$, the Planck length parameter by $\, \k = \sq{8 \pi G_N}$ 
(where $G_N$ is the Newton's gravitational constant) and the values of 
parameters or functions at the present epoch by an affixed subscript `$0$'.

\section{MST Cosmology in the Einstein frame, and the emergent DEM interacting 
scenario \label{sec:MST}}

As mentioned above, an intriguing scenario of an effective DEM interaction
emerges from a typical scalar-tensor equivalent MG formulation, viz. the one
involving a non-minimal metric-scalar-torsion (MST) coupling, in the Einstein 
frame
\cite{SSASB-MST}.
Let us first review briefly the main aspects of such a formalism, and the 
emergent cosmological scenario in the standard FRW framework.

Torsion, by definition, is a third rank tensor $T^\a_{~\m\n}$ which is 
anti-symmetric in two of its indies ($\m$ and $\n$), because of being the
resultant of the anti-symmetrization of a general asymmetric affine 
connection ($\Ct^\a_{\m\n} \neq \Ct^\a_{\n\m})$, that characterizes the 
four-dimensional Riemann-Cartan (or $U_4$) space-time geometry. The latter
however demands the metric-compatibility, viz. the condition $\, \nt_\a 
g_{\m\n} = 0$, where $\nt_\a$ is the $U_4$ covariant derivative defined in
terms of the corresponding connection $\Ct^\a_{\m\n}$. Such a condition 
leads to a lot of simplification in the expression for the $U_4$ curvature
scalar equivalent, $\Rt$, which is usually considered as the free $U_4$ 
Lagrangian in analogy with the free gravitational Lagrangian in GR, viz.
the Riemannian (or $R_4$) curvature scalar $R$. Specifically, $\Rt$ gets
reduced to a form given by $R\,$, plus four torsion-dependent terms 
proportional to the norms of irreducible modes, viz. the trace vector 
$\, \cT_\m \equiv T^\a_{~\m\a}$, the pseudo-trace vector $\, \cA^\m := 
\e^{\a\b\c\m} \,T_{\a\b\c} \,$ and the (pseudo-)tracefree tensor 
$\, \cQ^\a_{~\m\n} := T^\a_{~\m\n} + \frac 1 3 (\d^\a_\m \cT_\n - 
\d^\a_\n \cT_\m) - \frac 1 6 \e^\a_{~\m\n\s} \cA^\a \,$, as well as the 
covariant divergence of $\cT_\m$
\cite{shap-trev}.
In absence of sources (or the generators of the so-called {\it canonical 
spin density}), all the torsion terms drop out, and hence the $U_4$ 
theory effectively reduces to GR. The situation remains the same for
minimal couplings with scalar fields as well. However, such minimal 
couplings are themselves problematic, when it comes to assigning the
effective Lagrangian uniquely upon eliminating the boundary terms 
\cite{BOS-MST,BOS-QGbook,NFS-MST,FRM-MST,shap-trev}.
An easy cure is to resort to distinct non-minimal couplings of a given
scalar field $\f$, in general, with each of the constituent terms in $\Rt$
\cite{shap-trev}.
However, this implies the involvement of more than one arbitrary coupling 
parameters, which may affect the predictability and elegance of the theory.
Hence, it is much reasonable to consider a non-minimal $\f$-coupling with 
the entire $\Rt$, so that there is a unique (MST) coupling parameter (to
be denoted by $\b$, say)
\cite{SSASB-MST}.

Eliminating boundary terms, we obtain the auxiliary equation (or constraint) 
$\, \cT_\m = 3 \f^{-1} \pa_\m \f \,$, which implies that the (presumably 
primordial, and {\it a priori} massless) scalar field $\f$ acts as a source 
of the trace mode of torsion. Considering further, a mass term $m^2 \f^2$ 
induced by torsion's axial mode $\cA_\m$, via one of the possible ways 
mentioned above (in the Introduction), we get the effective MST 
action\footnote{Ignoring of course, any external source for the tensorial 
mode $\cQ^\a_{~\m\n}$, which therefore vanishes identically. This is 
particularly relevant to what we intend to study here, viz. a homogeneous 
and isotropic cosmological evolution in presence of torsion, which is 
plausible only when the latter's modes are severely constrained. One such 
constraint is $\cQ^\a_{~\m\n} = 0$
\cite{SSASB-tmax}.
} 
\cite{SSASB-MST}:
\be \label{J-MSTac}
\cS = \int d^4 x \sq{-g} \le[\fr{\b \f ^2} 2 R -\, \fr{1-6\b} 2 \, g^{\m\n} 
\pa_\m \f \, \pa_\n \f -\, \fr 1 2 \, m^2 \f ^2 +\, \cL^{(m)} \!\le(g_{\m\n},
\{\psi\}\ri)\ri] ,
\ee
which is nothing but the scalar-tensor action in presence of minimally 
coupled matter fields ($\{\psi\}$) described by the Lagrangian $\cL^{(m)}$, 
in the Jordan frame. 

Under a conformal transformation $\,g_{\m\n} \to \hg_{\m\n} \!= \!\le(\f/\fp\ri)^2 
\! g_{\m\n}\,$ and field redefinition $\, \vph := \fp \ln \le(\f/\fp\ri)\,$, with
$\, \fp = (\k \sq{\b})^{-1}$ --- the value of $\f$ at the present epoch $t = \tp$,
one obtains the Einstein frame MST action
\be \label{E-MSTac}
\hS = \int d^4 x \sq{-\hg} \le[\fr{\hR}{2 \k^2} -\, \fr 1 2 \, \hg^{\m\n}  
\pa_\m \vph \, \pa_\n \vph -\, \L e^{- 2 \vph/\fp} +\, \hLm \!\le(\hg_{\m\n},
\vph,\{\psi\}\ri)\ri] ,
\ee
where $\, \hR \,$ is the corresponding (Ricci) curvature scalar, and $\, \k = 
\sq{8 \pi G_N} \,$ denotes the gravitational coupling factor\footnote{This can
be actually be retrieved from the relationship $\, \k = (\f/\fp)\, \keff(\f) \,$, 
where $\, \keff(\f) = (\f \sq{\b})^{-1} \,$ is the effective (running) 
gravitational coupling one has in the Jordan frame.}. 
The parameter $\, \L = \fr 1 2 m^2 \fp^2\,$, which amounts to the effective 
field potential at $t = \tp$, and 
\be \label{E-matLag}
\hLm \!\le(\hg_{\m\n},\vph,\{\psi\}\ri) =\, e^{- 4 \vph/\fp} \, \cL^{(m)} 
\!\le(g_{\m\n},\{\psi\}\ri) ,
\ee
is the transformed matter Lagrangian, which depends on the field $\vph$ both 
explicitly as well as implicitly (since $g_{\m\n} = g_{\m\n} \!\le(\hg_{\m\n},
\vph\ri)$). It is in fact this $\vph$-dependence which leads to the DEM 
interaction in the standard cosmological setup, as we shall see below. Note 
also that, by definition, $\, \vph\big\vert_{t=\tp} \!= 0$.

Dropping the hats ($~ \widehat{} ~$), we express the gravitational field 
equation and the individual matter and field (non-)conservation relations 
in the Einstein frame as follows:
\bea 
&& R_{\m\n} -\, \fr 1 2 \, g_{\m\n} \, R =\, \k^2 \le[T_{\m\n}^{(m)} +\, 
T_{\m\n}^{(\vph)} \right] \,, 
\label{E-graveq} \\
&& \nab_\a \le(g^{\a\n} \, T_{\m\n}^{(m)}\ri) =\, -\, \nab_\a \le(g^{\a\n} 
\, T_{\m\n}^{(\vph)}\ri)   = -\, \fr{T ^{(m)} \pa_\m \vph}\fp \ , 
\label{E-consveq}
\eea
where $T_{\m\n}^{(m)}$ and $T_{\m\n}^{(\vph)}$ are the respective energy-momentum tensors for matter and scalar field
and $\, T^{(m)} \equiv\, g^{\m\n} \, T_{\m\n}^{(m)} \,$ denotes the trace of
$ T_{\m\n}^{(m)}$.

Considering the matter to be {\it a priori} in the form of a pressure-less 
fluid (viz. `dust'), we have in the standard spatially flat FRW framework, 
$\, T^{\m \,(m)}_{~\n} =\, \text{diag} \le[- \rmt, \,0, \,0, \,0\ri]$, so 
that $\, - T^{(m)} = \rmt\,$ is just the matter density, which is purely 
a function of the cosmic time $t$. 
Because of the interaction (\ref{E-consveq}),
the matter density ceases to have its usual dust-like evolution (i.e. $\rmt(t)
\not\propto a^{-3} (t)$, where $a(t)$ is the FRW scale factor). Nevertheless, 
the above Eqs.\,(\ref{E-graveq}) and (\ref{E-consveq}) are shown to be solvable 
in an exact analytic way, for the configuration
\be
\vph(t) =\, s \, \fp \, \ln\le[a(t)\ri] \,, \qquad \rmt(t) \propto a^{-(3+s)} 
(t) \,,
\ee
provided one sets the constant parameter $\, s = 2\b$
\cite{SSASB-MST}.
Consequently, the matter density parameter $\Om(a)$ is expressed as
\be {\label{MDP}}
\Om(a) :=\, \fr{\rmt(a)} {\r(a)} =\, \fr{ (3-s) \, \Omp a^{-(3-s)} } {3\Omp (a^{-(3-s)}-1) 
+ (3-s)} \ ,
\ee
where $\r(a)$ is the total (or critical) density of the universe and $\Omp$ is the value of 
$\Om$ at the present epoch ($t = \tp$, whence $a = 1$). Using the Friedmann and Raychaudhuri 
equations we can then express the Hubble 
parameter and total EoS parameter of the system, respectively, as
\be
H(a) :=\, \fr{\dot a} a =\, \Hp \left(1-\fr{s}3\right)^{-1/2}\left[\Omp  a^{-(3+s)} + \left(1-\fr{s}3 - \Omp 
\right)a^{-2s}\right]^{1/2}, 
\ee
\be {\label{EOS}}
\sw(a) := \fr{p(a)} {\r(a)}= - 1  + \Om(a) + \fr{2s} 3 \ ,
\ee
where $\Hp = H (a=1)\,$ is the Hubble constant, and $p (a)$ denotes the total pressure. 
Note that in the limit $s \to 0$, the above equations reduce to the corresponding ones 
for $\L$CDM. Therefore one can directly estimate the extent to which the MST 
cosmological scenario can deviate from $\L$CDM, by demanding that such a deviation
should not breach the corresponding $68\%$ parametric margins for $\L$CDM. This would
in turn provide an estimation of the parameter $s$, which has actually been carried 
out in 
\cite{SSASB-MST},
using the Planck 2015 and the WMAP 9 year results. The upper bound on $s$, thus obtained,
is of the order of $10^{-2}$. Nevertheless, a rather robust reconciliation is required
from an independent analysis, for instance, using the RSD and $H(z)$ observations, which
we endeavor to do in this paper. 

\section{Growth of matter density perturbations} {\label{sec:MPGrowth}}

In this section, we discuss the evolution of linear matter density perturbations in the 
deep sub-horizon regime for the aforementioned Einstein frame background MST cosmological
scenario. The perturbations can be studied in the well-defined conformal Newtonian gauge. 
The metric in this gauge is given as
\cite{AT-book}
\be \label{pert-metric}
ds^2 = e^{2N}[-(1-2\F)\cH ^{-2} dN^2 + (1+ 2\F)\d_{ij} dx^i dx^j] \ ,
\ee
where $N:=\ln a(t)$ is the number of e-foldings, $\cH$ is the conformal Hubble parameter 
and $\F$ is the Bardeen potential. Note that we have taken the same potential $\F$ in both
temporal and spatial part of the metric under the assumption of a vanishing anisotropic 
stress.

The evolution of the matter density contrast $\dmt$ depends on the divergence or convergence 
of the peculiar velocity $\vec{\pmb v}^{(m)}$ via the perturbed continuity equation
\be {\label{dprime}}
\fr{d \dmt} {d N}  = -\tmt \ , \quad \mbox{where} \quad  
\tmt :=\nabla \cdot \vec{\pmb v}^{(m)} \,.
\ee
On the other hand, the \textit{Euler} equation for matter perturbations is given by 
\be {\label{thprime}}
\fr{d \tmt}{d N} =  -\le[\fr{\tmt} 2 \le(1-3\,\sw -  \k \sq{2 s} \,  \fr{d \vph}{d N} \ri) + 
\hl ^{-2} \le(\F + \k \sq{\fr{s} 2} \, \d\vph \ri)\ri] \,,
\ee
where $\hl \equiv \cH/ k$ (with $k$ being the comoving wavenumber), and 
\be
\F \simeq\, \fr{3} 2 \hl^{2}\Om \dmt \ , \quad  \mbox{and} \quad <\!\d \vph\!> \,\simeq\, 3 \hl^2\sq{\fr{s} 2}\,\Om \dmt \,,
\ee 
considering only the mean value of $\d \vph$, as it shows a damped oscillatory behavior in
the sub-horizon regime.

$\F$ and $\d \vph$ both being proportional to $\hl^2$, become negligible in the deep 
sub-horizon limit ($\hl^{2} \ll 1$). However, their contribution may not be negligible 
in the evolution of $\tmt(N)$, because of the $\,\hl ^{-2}$ pre-factor in the second term
of Eq.\,(\ref{thprime}). As a consequence, the DE perturbation $\d \vph$ which itself is 
negligible in the sub-horizon regime (despite being scale-dependent) may, by virtue of 
its coupling with matter, lead to a significant effect on the growth of matter density 
perturbations.

\begin{figure}[htb] 
\centering
\begin{subfigure}{0.495\linewidth} \centering 
   \includegraphics[scale=0.84]{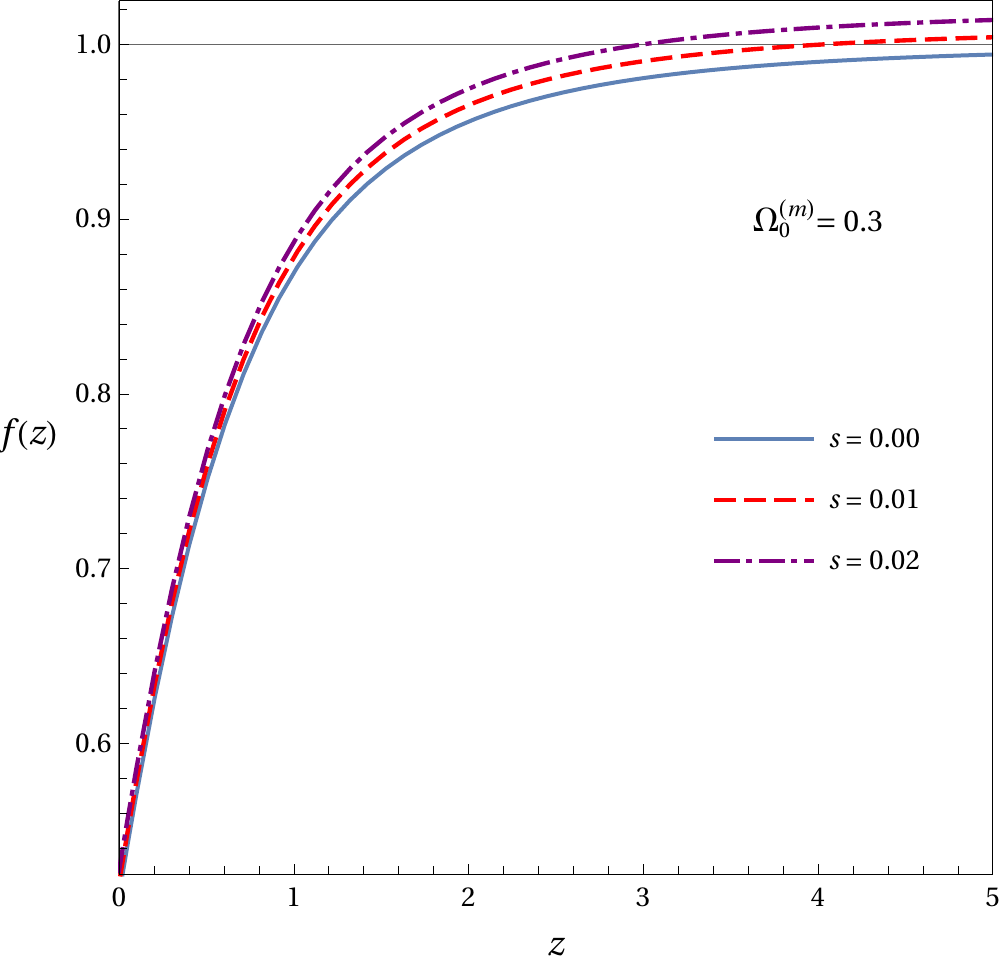}
   \caption{\footnotesize Growth factor evolution for fiducial $\Omp$ and $s$.} 
   \label{f-fig}
\end{subfigure}
\begin{subfigure}{0.495\linewidth} \centering
    \includegraphics[scale=0.88]{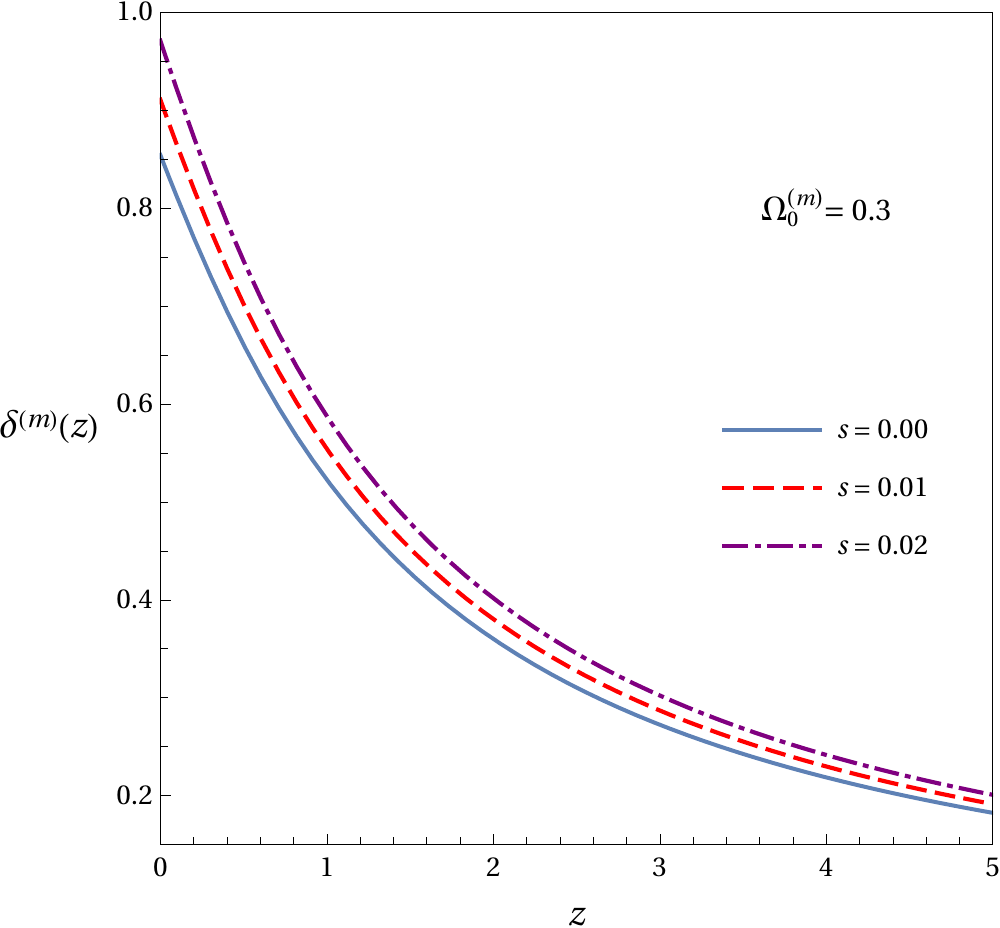}
    \caption{\footnotesize Density contrast evolution for fiducial $\Omp$ and $s$.} 
    \label{delta-fig}
\end{subfigure}
\caption{\footnotesize Functional variations of the growth factor and the matter density
contrast, $f(z)$ and $\dmt(z)$ respectively, in the redshift range $z\in[0,5]$, for 
certain fiducial parametric settings, viz. $\Omp = 0.3$ (fixed) and $s = 0$, $0.01$, and
$0.02$.}
\label{fig1}
\end{figure}

Eqs.\,(\ref{EOS}), (\ref{dprime}) and (\ref{thprime}) yield the second-order differential
equation
\be {\label{DPE}}
\fr{d \dmt}{d N} + \le[2(1-s)-\fr{3 \Om} 2 \ri] \fr{d \dmt}{d N} =\, 
\fr {3 (1+s)} 2 \, \Om \dmt \,,
\ee  
which can be reduced to the following first-order differential equation:
\be {\label{GFE}}
\fr{d f}{d N} + f^2 + \le[2(1-s) - \fr{3 \Om} 2\ri] f = \fr{3 (1 + s)} 2 \, \Om \,,
\ee
by defining the so-called growth factor $\, f(N) := d[\ln{\dmt}]/d N$
\cite{GAC-mp,BGNP-NS,CFGPS-Ggi,KD-Rg,TGMP-mpfR}. 
Due to the pre-factor $(1+s)$ in the r.h.s. of Eq.\,(\ref{GFE}), the function $f(N)$ can 
cross the unity barrier at high redshifts (whence $\Om \to 1$). This is illustrated in 
Fig. (\ref{f-fig}), where we have plotted $f(z)$ for a fixed $\Omp = 0.3$ and certain 
fiducial values of $s$, including $s=0$ (the $\L$CDM case). Fig. (\ref{delta-fig}), on the 
other hand, depicts the evolution of $\dmt(z)$, which tends to increase with $s$ for a 
fixed $\Omp = 0.3$.
%
%%%%%%%%%%%%%%%%%%%%%%%% figure 1 %%%%%%%%%%%%%%%%%%%%%%%%
%%%%%%%%%%%%%%%%%%% figure 1 ends %%%%%%%%%%%%%%%%%%%%%%%%%
% 

\subsection{Growth factor parametrization \label{sec:GF}}

As mentioned earlier, following the well-known prescription of 
\cite{PA-grow,PG-grow} 
we may consider parametrizing the growth factor $f(z)$ as $[\Om(z)]^{\c(z)}$. However, such 
a parametrization does not explain the crossing of $f(z)$ from $<1$ to $>1$ at large-redshifts, 
as illustrated in Fig.\,(\ref{f-fig}). More precisely, this parametrization $f(z)$ always 
restricted within the range $[0,1]$ at all redshifts which in our case is not true. 
So to alleviate this limitation, we propose the ansatz:
\be \label{GF}
f(z) =\, (1+s) \le[\Om(z)\ri]^{\c(z)}  \,,
\ee
which evidently implies $f(z)$ approaching $1+s$ at large redshifts (whence $\Om \to 1$). 
Now, to determine the growth rate of matter perturbations from Eq.\,($\ref{GF}$), it is 
necessary to find a suitable functional form of $\c(z)$. In particular, choosing to express 
the growth index as a function of the scale factor $a$, we in this paper resort to the
following truncated form of its Taylor expansion about $a=1$ (which corresponds to the 
present epoch): 
\be \label{GI}
\c(a) =\, \cp +\, \cpp\, (1 - a) \,, \qquad \mbox{with \,\,$\cp, \cpp :=$ constants} \,,
\ee
as in
\cite{WYF-grwpara,BA-grwindx}.
Note that this parametrization is valid atleast upto a redshift $z \simeq 1$ and is 
therefore suitable for the analysis using the RSD observational dataset
\cite{beut-6df,ata-dr14,beut-boss},
as most of the 
data points in that set lie within $z=1$. In fact, it is rather convenient for us to 
re-write Eq.\,(\ref{GI}) as
\be \label{g}
\c(N) = \cp + \cpp (1-e^N) \quad \mbox{with} \quad \cp = \c(N)\big|_{N=0} \,, \quad 
\cpp = \fr{d\c(N)} {dN}\Big |_{N=0} \,,
\ee
where
\bea 
\cp &=& \c(N)\big|_{N=0} =\, \fr 1 {\ln \Omp} \ln \le(\fr{\gfp}{1+s}\ri) \, \qquad 
\le[\gfp = f|_{_{N=0}}\ri] \, , 
\label{g0} \\
\cpp &=& \fr{d\c(N)} {dN}\Big |_{N=0} =\, \fr 1 {\ln \Omp } \Big[\cp (s-3+3\Omp) + (1+s)(\Omp)^{\cp} \nn\\
&& \hspace{100pt} +\, 2(1-s) - \fr{3} 2 \Big(\Omp + (\Omp)^{1-\c _0}\Big) \Big]  \, ,
\label{g'}
\eea
by Eqs.\,(\ref{GFE}) and (\ref{GF}). 

For the $\L$CDM case ($s=0$), assuming $\Omp=0.3$, one gets $\cp \simeq 0.555$ and
$\cpp \simeq -0.016$. Moreover, the signature of $\cpp$ can discriminate between 
various DE models and modified gravity theories. For instance, the minimal level
Dvali-Gabadadze-Porrati (DGP) model predicts $(0.035 \!< \!\cpp \!< \!0.042)$ 
which is in sharp contrast with the GR predictions 
\cite{WYF-grwpara}.
%

%%%%%%%%%%%%%%%%%%%%%%%%%%%%%%%%%%%%%%%%%%%%%%%%%%%%%%%%%%%%
\subsection{Numerical fitting of growth index} {\label{sec:GIFit}}

Let us now focus on determining the parametric set $p(\th)=\{s,\Omp,\sep,\cp,\cpp\}$. 
While the form of the parameter $\cpp$ is already obtained in terms of $s$, $\Omp$ and 
$\cp$, we require to assert the form of $\cp$ in the first place. However, as we see
from Eq.\,(\ref{g0}), $\cp$ depends on $s$ and $\Omp$ as well. Hence we resort to
solving numerically the differential equation (\ref{GFE}), by taking $s \in [0,0.1]$ 
and $\Omp \in [0.2,0.4]$ (which are of course fairly wide range of values), and for a 
step-size of $0.01$. Using Eq.\,(\ref{g0}) thereafter, we obtain the following fit:
\be \label{gamma0-fit}
\cp \simeq\, \fr{0.547}{[\Omp]^{0.012}}\, -\, 1.118 \, s \, \Omp  \ .
\ee
%
%%%%%%%%%%%%%%%%%%%%%%%% figure 2 %%%%%%%%%%%%%%%%%%%%%%%%
\begin{figure}[htb]
\centering
\begin{subfigure}{0.495\linewidth} \centering
   \includegraphics[scale=0.84]{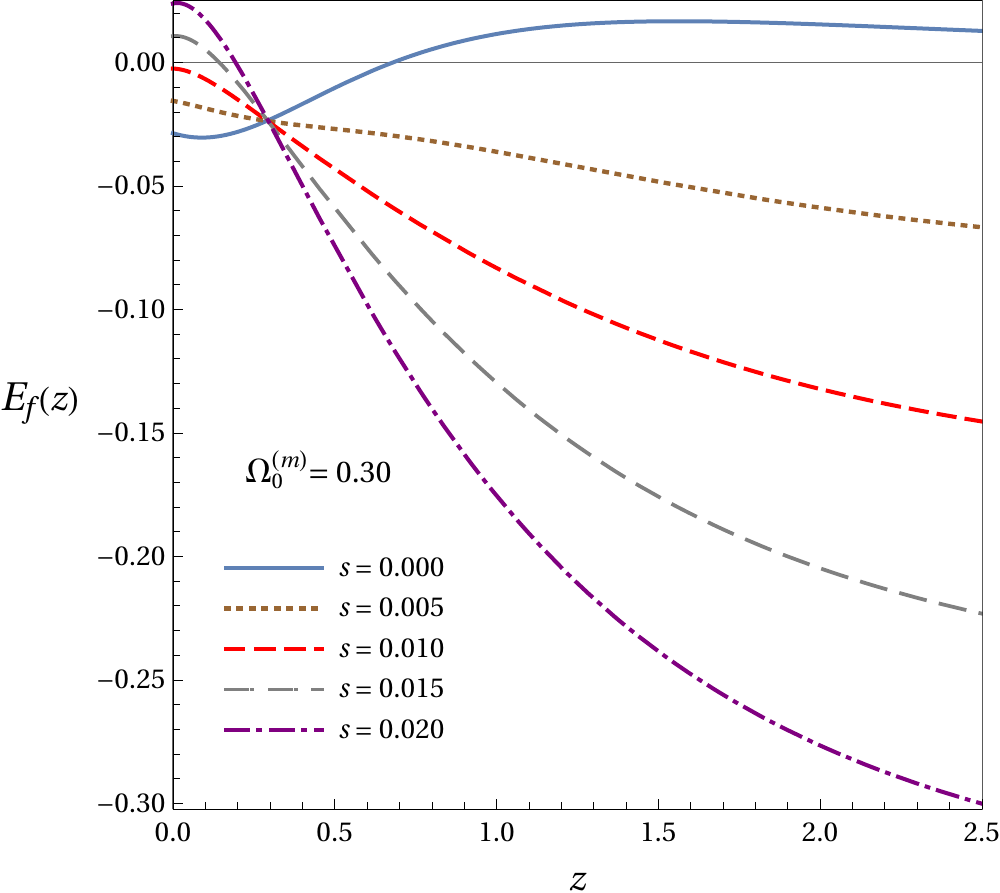}
   \caption{\footnotesize Growth factor fitting error for fixed $\Omp$ and variable $s$.} \label{ferror(a)}
\end{subfigure} 
\begin{subfigure}{0.495\linewidth} \centering
    \includegraphics[scale=0.84]{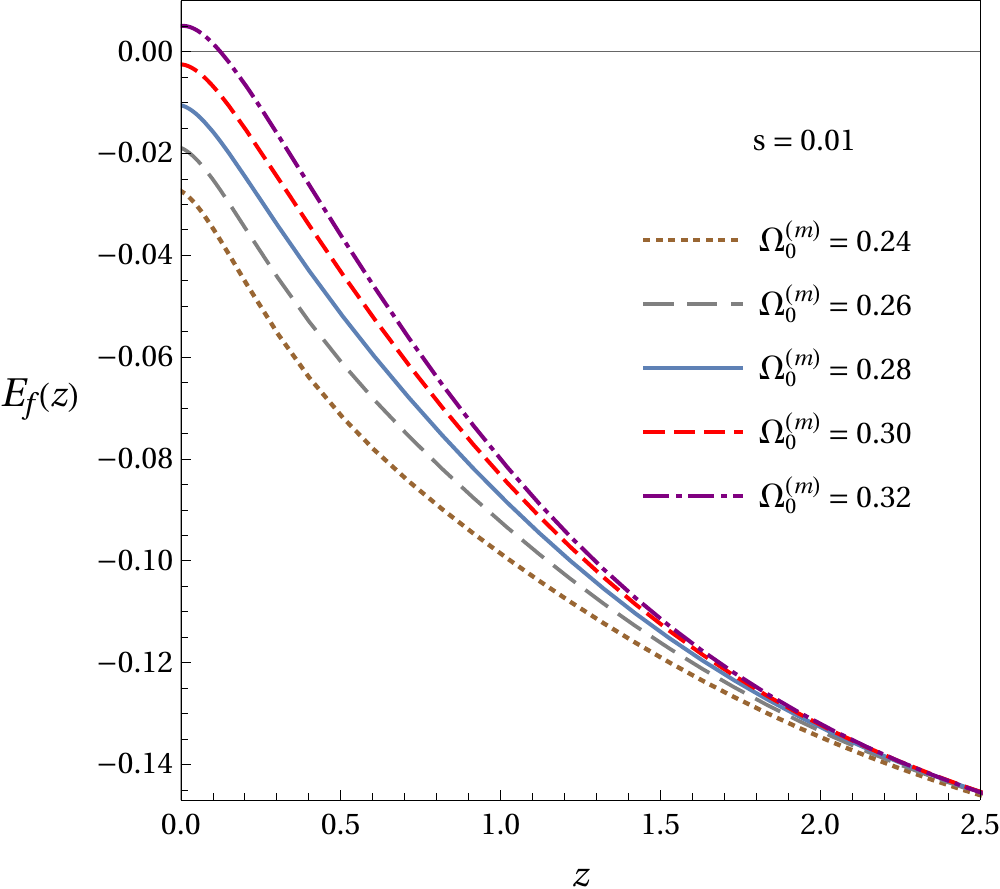}
    \caption{\footnotesize Growth factor fitting error for fixed  $s$ and variable $\Omp$.} \label{ferror(b)}
\end{subfigure} 
\caption{\footnotesize Functional variations of the growth factor fitting error, $E_f(z)$,
in the redshift range $z\in[0,2.5]$, for certain fiducial parametric settings.}
\label{fig2}
\end{figure}
%%%%%%%%%%%%%%%%%%% figure 2 ends %%%%%%%%%%%%%%%%%%%%%%%%%
%

\noindent
In order to verify the validity of this fitting, let us take the $\Omp = 0.3$, say, and 
the limit $s \rightarrow 0$. Eq.\,(\ref{gamma0-fit}) then gives $\cp \simeq 0.555$ which 
is precisely what we had estimated theoretically, for the $\L$CDM case, in the last 
subsection, by using Eqs.\,(\ref{GFE}) and (\ref{g0}). The goodness of the fit is 
illustrated in Figs.\,(\ref{ferror(a)}) and (\ref{ferror(b)}), in which we have plotted 
the fractional error in the fitting, viz. $E_f(z) (\,= [f_F(z) - f(z)]/f(z)\,$ with 
$z \in [0,2.5]$, for a fixed $\Omp = 0.3$ and a range of fiducial values of $s$, and for
a fixed $s = 0.01$ and a range of fiducial values of $\Omp$, respectively. In both the
cases, the error turns out to be $\simeq 0.2\%$ at $z \simeq 1$, indicating a fair amount
of the accuracy of the fit.

\section{Parametric estimations from RSD and Hubble observations} {\label{sec:RSDEst}}
   
After formulating $\cp$ and $\cpp$ in terms of $s$ and $\Omp$, we are left with only 
three parameters $s$, $\Omp$ and $\sep$ in hand. So in order to estimate them from the observations use the $f\se(z)$ observations from various galaxy data 
surveys 
\cite{beut-6df,ata-dr14,beut-boss,sanc-boss,FNB-Gcp,blake-GAMA,SP-sf,shi-rss,howl-MTF,torr-VIP,marin-sdss}, 
we will now proceed to perform the statistical analysis, in particular MCMC 
simulation to estimate our model parameters. 
Theoretically, $(f\se)_{th}(z)$ can be written as 
\cite{SPR-rss,KPS-brs,KP-tens,DL-tDE,HSK-mic,NPP-tens}
\be \label{gr}
(f\se)_{th}(z) = f(z)\sep \fr{\dmt(z)} {\dmp} \ , \quad \mbox{where} \quad \sep = \se|_{z=z_0} \ .
\ee
which can be explicitly written as
\be
(f\se)_{th}(N) =  \sep (1+s) ({\Om})^{\c(N)} \exp \left[(1+s) \int ^N_0 {(\Om})^{\c(N)} dN \right] \ ,
\ee
where we have used Eq. (\ref{GF}). Since our parameter $s$ is presumably positive definite and small, it is convenient for us to write $s = |\st|$, where $\st$ can take both positive and
negative values.

In order to perform the standard $\chi^2$ minimization, we use the growth data observations: 
$A_{obs} \equiv (f\se)_{obs}$ along with the theoretical predicted values: $A_{th} \equiv (f\se)_{th}$ 
in the standard definition of the $\chi^2$ function
\be {\label{chisquare}}
\chi^2 : = V^m C_{m n}^{-1} V^n, 
\ee
where $V := A_{obs}-A_{th}$ and $C_{m n}^{-1}$ is the inverse of the covariance matrix 
between three WiggleZ data points 
\cite{KP-tens}.
As we have already shown in fig. (\ref{fig2}) 
that the parametric form (\ref{g}) tends to diverge in case of interacting DE from its numerical solution (\ref{GF}) 
at high redshifts, we therefore restrict ourselves for the observations upto $z=1$ for the datasts: GOLD-2017
\cite{NPP-tens} ,
and
$H(z)$ data set 
\cite{RDR-bao}.
%
%%%%%%%%%%%%%%%%%%%%%%%% figure 3 %%%%%%%%%%%%%%%%%%%%%%%%
\begin{figure}[!htp]
\centering
\begin{subfigure}{0.485\linewidth} \centering
   \includegraphics[scale=0.45]{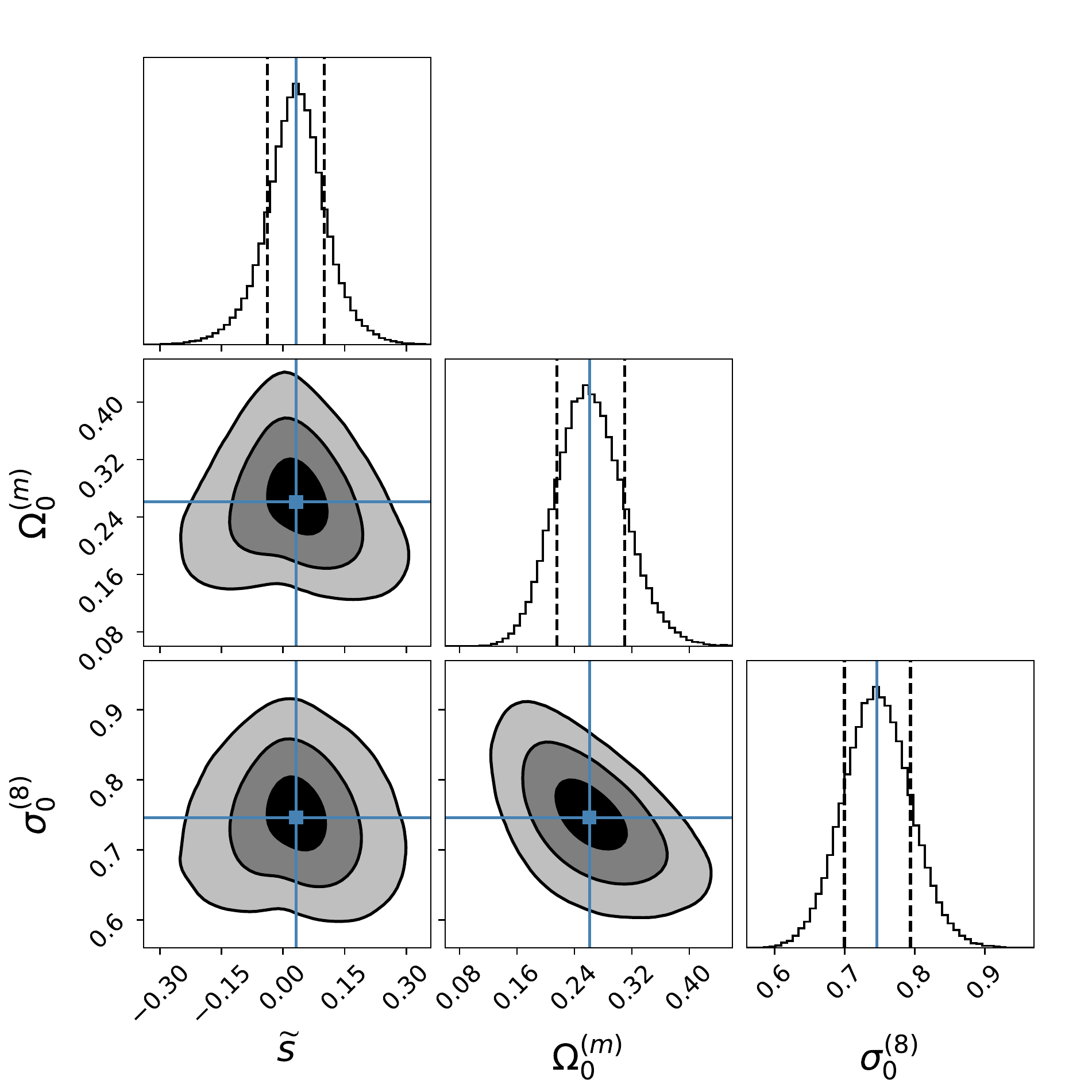}
   \caption{\footnotesize $2$-D posterior distribution for Gold data.}
   \label{Gold}
\end{subfigure} 

\begin{subfigure}{0.485\linewidth} \centering
    \includegraphics[scale=0.45]{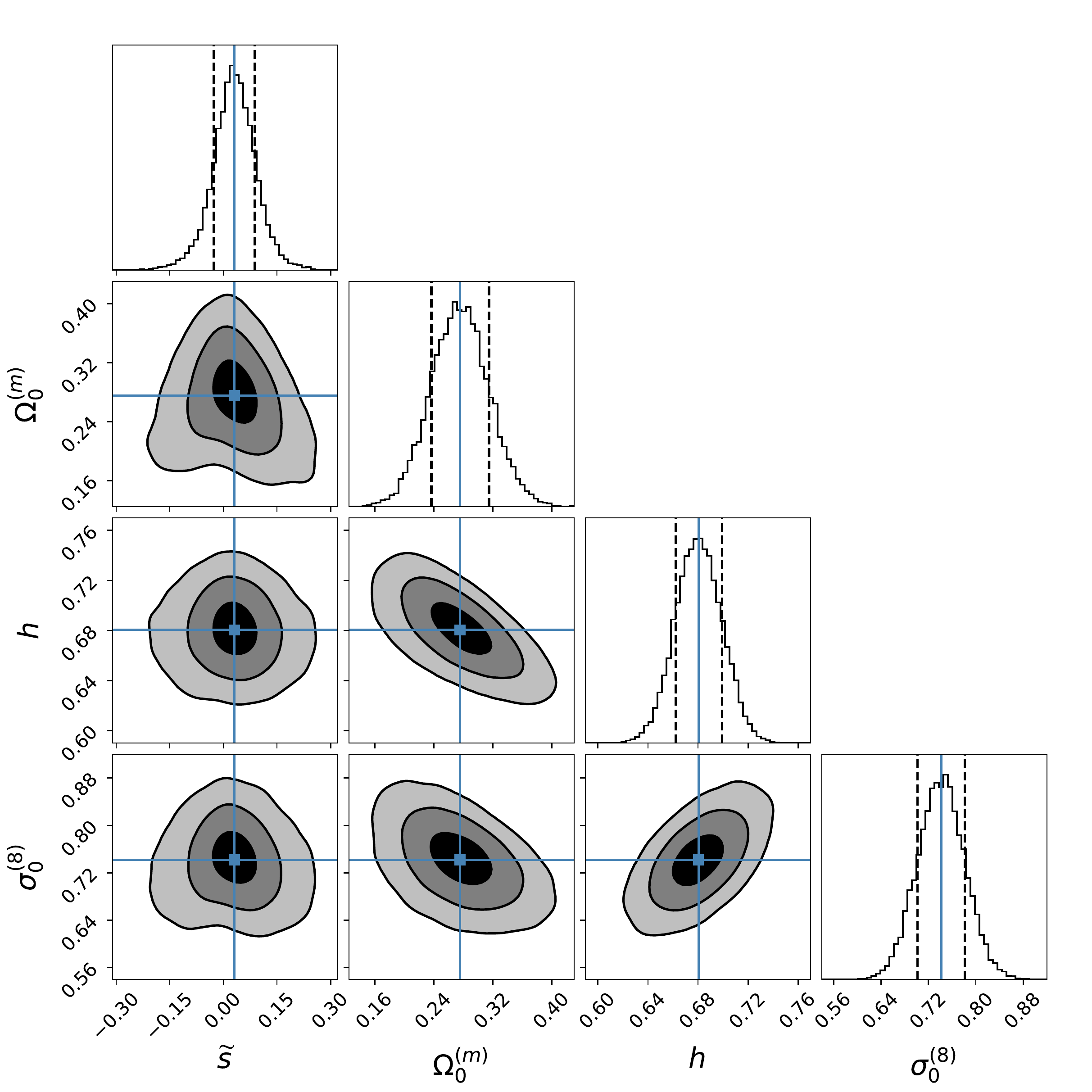}
    \caption{\footnotesize $2$-D posterior distribution for GOLD$+ H(z)$ data.}
    \label{GoldHubb}
\end{subfigure} 
\caption{\footnotesize The $1\s$-$3\s$ contour levels for Gold
dataset (left), and its combination with the Hubble dataset (right). The solid blue line denotes the best-fit and dashed lines correspond to the $1\s$ level. }
\label{fig3}
\end{figure}
%%%%%%%%%%%%%%%%%%% figure 3 ends %%%%%%%%%%%%%%%%%%%%%%%%%
%
Also, we set the range of priors as follows: 
($i$) $-1 \leq \st \leq 1$, ($ii$) $0.1 \leq \Omp \leq 0.6$, ($iii$) $0.5 \leq \sep \leq 1.2$ and 
($iv$) $0.4 \leq h \leq 0.9$, where $h:= \Hp/[100 \, Km \, s^{-1}\, Mpc^{-1}]$. The obtained 
contour plots between parameters upto $3\s$ level are shown in Figs. \ref{Gold} and \ref{GoldHubb}.
%
%%%%%%%%%%%%%%%%%%%%%%%%%%

\begin{table*}[ht]
\centering
{
\begin{tabular}{||p{3cm}||p{2.3cm}|p{2.3cm}|p{2.3cm}|p{2.3cm}||p{1.5cm}||}
\hline
 & \multicolumn{4}{c||}{Parametric estimations}
& \\
\hspace{0.1in} Observational  &  \multicolumn{4}{c||}{\footnotesize (best fit \& $68\%$ limits)}  & ~ $\chi ^2/dof $ \\
\cline{2-5}
%\cline{7-8}
%
\hspace{0.35in} datasets & \hspace{0.33in} $\Omp$ &  \hspace{0.33in} $\sep$ 
 &  \hspace{0.33in} $h$ & \hspace{0.33in} $\st$ &  \\
\hline\hline
1. {\small GOLD} & \hspace{0.1in}$ 0.2610^{+0.0487}_{-0.0451}$ & \hspace{0.1in}$0.7460^{+0.0476}_{-0.0466}$  & \hspace{0.4in}- & \hspace{0.1in}$0.0319^{+0.0683}_{-0.0702}$ & \hspace{0.05in} $0.8390$  \\
\hline
2. {\small GOLD$+ H(z)$ } & \hspace{0.1in}$0.2753^{+0.0393}_{-0.0387}$ & \hspace{0.1in}$ 0.7417^{+0.0393}_{-0.0402}$  & \hspace{0.1in}$0.6804^{+0.0189}_{-0.0183}$  & \hspace{0.1in}$ 0.0308^{+0.0573}_{-0.0573}$ & \hspace{0.05in} $0.6125$ \\
\hline\hline
\end{tabular}
}
\caption{\footnotesize Best fit values with $1\s$ confidence limits of parameters 
$\Omp$, $\sep$, $h$ and $\st$ together with their corresponding $\chi ^2/dof$ for GOLD and GOLD $+H(z)$ dataset.}
{\label{E-tab1}}
\end{table*} 

The estimations are shown in table (\ref{E-tab1}) in which one can see that the best-fit of $\st$ for both sets of data (GOLD and GOLD$+ H(z)$) is insignificant (as expected, since observations mostly prefer the $\L$CDM model), but even then within $1\s$ limits its domain can reach upto significantly large value i.e. $\cO(10^{-2})$ which shows a reasonable large deviation from the $\L$CDM model. This indicates from the low-redshift data we can still observe a convincing amount of DEM interaction even at the $1\s$ level.

%%%%%%%%%%%%%%%%%%%Conclusion%%%%%%%%%%%%%%%%%%%%%%%%
\section{Conclusion \label{sec:Concl}}

We have formulated the growth of linear matter density perturbations in a parametric form for a DE model which stems out from a modified gravity approach consists of metric and torsion as two basic entities of the space-time geometry. In the formalism, we have briefly demonstrated that a non-minimal coupling of metric and torsion with scalar field can give rise to a scalar-tensor action of DE in the Jordan frame which upon conformal transformation to the Einstein frame naturally makes scalar field non-minimally coupled with the matter sector. Due to this coupling, matter and scalar field exchange their energies between each other which ceases their individual energy densities to be self-conserved. The latter, thus, has direct influence on the underline matter density contrast and its evolution, which we have explored in this work.

We have demonstrated that in the perturbed FRW space-time, the scalar field and matter coupling enhances the growth of matter density perturbations in the sub-horizon regime, allowing it to cross the upper barrier of unity at large redshifts. Since this effect is unique in the interacting DEM scenarios it requires a slight modifications in the standard parametric ansatz of growth factor. With suitable modification we propose a slightly different growth factor ansatz to make the parametric formulation compatible with the theoretical predictions. Also, in view of the time evolving growth index, which is even encountered for the $\L$CDM model, we have chosen an appropriate functional form i.e. first order Taylor expansion about present-day value of the scale factor $a(t)$. This simple but well defining form of the growth index indeed illustrates the parametric formulation of growth factor close to its actual evolution atleast upto $z \simeq 1$. Since the present-day value of growth index itself depends on the background model parameters, therefore in order to choose its explicit function form we have numerically obtained its fitting formula in terms of coupling as well as energy density parameter which we have shown to be a well approximation for a wide range of parameters. 

As to the parametric estimations, we have constrained parameters $\st$, $\Omp$ and $\sep$ by 
using the RSD as well as its combination with the Hubble data. We have found that for the GOLD datset the $\st$ and hence $s$ parameter can show mildly large deviation from the $\L$CDM model upto $1\s$, which is comparatively smaller for the combined dataset, as expected. The consistency in our estimations with the theoretical predictions confirms the validity of our fitting function. However, to explain growth history for redshifts $>1$, the above parametrization still requires further modifications to deal with various DE models as well as modified gravity theories, which we will shall endeavor to report in near future.

\section{Acknowledgments}
The work of MKS was supported by the Council of Scientific and Industrial Research (CSIR), Government of India. 

\end{document}